<u>Title</u>: Fire responses shape plant communities in a minimal model for fire ecosystems across the world

<u>Short title</u>: Fire responses and plant communities

<u>Keywords</u>: fires, plant fire response, plant traits, plant communities, alternative ecological states, ecological modeling

<u>Authors</u>: Marta Magnani[1,2,3*], Rubén Díaz-Sierra[4,5], Luke Sweeney[6,7], Antonello Provenzale[1,9], Mara Baudena[8,9,3,4]

\* <u>Corresponding author, e-mail</u>: Marta Magnani, marta.magnani@igg.cnr.it

<u>Affiliations</u>:

[1] Institute of Geoscience and Earth Resources, CNR, Via Valperga Caluso, 35, 10125 Torino, Italy

[2] Università degli studi di Torino and INFN, Via P. Giuria 1, 10125 Torino, Italy

[3] Centre of Complex System Studies (CCSS), Utrecht University, 3508 TC Utrecht, The Netherland

[4] Copernicus Institute of Sustainable Development, Utrecht University, 3508 TC Utrecht, The Netherland

[5] Universidad Nacional de Educación a Distancia, UNED, Madrid, Spain

[6] School of Archaeology, Geography and Environmental Sciences (SAGES), University of Reading, UK

[7] Leverhulme Centre for Wildfires, Environment and Society, Imperial College London, South Kensington, London, SW7 2BW, UK

[8] National Research Council of Italy, Institute of Atmospheric Sciences and Climate (CNR-ISAC), CNR, Corso Fiume 4, 10133, Torino, Italy

[9] National Biodiversity Future Center, Palermo 90133, Italy

Emails:

Marta Magnani: marta.magnani@igg.cnr.it

Rubén Díaz-Sierra: sierra@ccia.uned.es

Luke Sweeney:  l.sweeney@pgr.reading.ac.uk

Antonello Provenzale: antonello.provenzale@cnr.it

Mara Baudena: m.baudena@isac.cnr.it

ORCIDs:

Marta Magnani: https://orcid.org/0000-0003-3180-1321

Rubén Díaz-Sierra: https://orcid.org/0000-0001-9821-8347

Luke Sweeney: https://orcid.org/0000-0002-8161-819X

Antonello Provenzale: https://orcid.org/0000-0003-0882-5261

Mara Baudena: https://orcid.org/0000-0002-6873-6466

# Abstract

Across plant communities worldwide, fire regimes reflect a combination of climatic factors and plant characteristics. To shed new light on the complex relationships between plant characteristics and fire regimes, we developed a new conceptual, mechanistic model that includes plant competition, stochastic fires, and fire-vegetation feedback. Considering a single standing plant functional type, we observed that highly flammable and slowly colonizing plants can persist only when they have a strong fire response, while fast colonizing and less flammable plants can display a larger range of fire responses. At the community level, the fire response of the strongest competitor determines the existence of alternative ecological states, i.e. different plant communities, under the same environmental conditions. Specifically, when the strongest competitor had a very strong fire response, such as in Mediterranean forests, only one ecological state could be achieved. Conversely, when the strongest competitor was poorly fire-adapted, alternative ecological states emerged, for example between tropical humid savannas and forests, or between different types of boreal forests. These findings underline the importance of including the plant fire response when modeling fire ecosystems, e.g. to predict the vegetation response to invasive species or to climate change.

# Introduction

Understanding the complex relationships between fire and its drivers is essential for both predicting environmental change in fire-prone biomes and assisting in fire management practices. Climatic drivers are generally used to predict fire frequency, fire season and burned area (Westerling and Bryant 2008; Jolly et al. 2015; Abatzoglou and Williams 2016; Boer et al. 2016; Ruffault et al. 2017; Turco et al. 2017, 2018). On the other hand, fire regimes, which include the type, frequency, intensity, seasonality and spread of recurrent fires (Gill 1975; Turner et al. 1998; Turner 2010), also depend on biological feedbacks (Thom and Seidl 2016; Pausas and Ribeiro 2017; Archibald et al. 2018; Pausas and Keeley 2019). Plant types influence fire primarily in terms of the availability, continuity and flammability of fuel (Wells et al. 2004; Bowman et al. 2009; Higuera et al. 2009; Karavani et al. 2018). Within each climate zone, plant characteristics can help to explain the occurrence of different fire regimes in different biomes (Archibald et al. 2013; Pausas and Ribeiro 2013). In tropical ecosystems, for instance, fast-growing and drying savanna grasses encourage frequent and low intensity fires, thus preventing the growth of forest trees that are poorly adapted to fires (Beckage et al. 2011; Ratnam et al. 2011). Such vegetation-fire feedback has been suggested to preserve savannas in areas where a closed humid forest might be expected based on climatic conditions alone (Van Langevelde et al. 2003; Bond 2008; Dantas et al. 2016; D'Onofrio et al. 2018). Similar examples of fire's role in maintaining ecological stability have been shown in boreal (Johnstone et al. 2010; Rogers et al. 2015; Couillard et al. 2018; Abis and Brovkin 2019) and temperate forests (Kitzberger et al. 2012, 2016; Tepley et al. 2016). Thus, fire regimes involve several feedbacks between plants, fires and climate, at differing spatial and temporal scales (Wright and Clarke 2007; Ali et al. 2008; Pausas and Keeley 2009; Johnstone et al. 2010; Archibald et al. 2018; Karavani et al. 2018).

In fire-prone environments, plant communities are shaped both by community dynamics, such as competition, and by fires (Lavorel and Garnier 2002). These factors are reflected in plant traits (Reich et al. 2003), including multiple types of plant adaptations to the local fire regime (Keeley 1986; Gignoux et al. 1997; Keeley et al. 2011).The traits that allow a species to survive within a particular environment are often correlated, creating so-called trait "syndromes" (Reich et al. 2003; Archibald et al. 2018).  In addition, for a given species, plant traits and trait syndromes reflect the trade-off between strategies (Grime 1977; Chapin III et al. 1993). For instance, tropical forest trees invest resources in fast growth between fire events (Rossatto et al. 2009; Viani et al. 2011) rather than investing in individual plant protection from fire damages (Hoffmann et al. 2012; de L. Dantas et al. 2013).

Among fire-adapted species, three fire syndromes can be identified (Pausas 2015*a*). These correspond to species that survive fires either at individual or at population level, or to species that do not tolerate fires. Plants that cope with fire at individual level (also named 'fire resisters', or 'fire survivors' in Schwilk and Ackerly 2001) may have thicker bark, which limits the damage to the tree during relatively low intensity, surface fires (Keeley et al. 2011), or may readily re-sprout after intense fires thanks to large below-ground carbohydrate reserves (Gignoux et al. 1997; Bond and Midgley 2001). Species that survive fires at population level (also called 'fire embracers') generally have elements of their life cycle closely tied to fire, including germination caused by combustion, post-fire seed release in crown systems (serotiny) or enhanced flammability to increase the frequency and intensity of fires to the detriment of non-resprouting competitors (Schwilk and Ackerly 2001; Keeley et al. 2011). Finally, fire-intolerant species (or 'fire avoiders') may have few adaptations to fire and are generally found in areas where fires are infrequent (Pausas 2015*a*).

In this study, we use a newly developed conceptual model (*sensu* Robinson 2008*a*, 2008*b*) to investigate the emergence of different plant communities in consequence of plant-fire interactions and plant competition for resources. We classified plants into functional types (PFTs), defined in terms of plant structure, response and functioning that are related to different sets of traits (Box 1996; Pausas and Lavorel 2003; Lavorel et al. 2007). This minimal model was a convenient framework for highlighting the general conceptual relationships between fires, plant characteristics and community composition. Similar approaches have been developed for specific biomes, such as savannas (Beckage et al. 2009, 2011; Baudena et al. 2010; De Michele et al. 2011; Ratajczak et al. 2011; Staver and Levin 2012), the Mediterranean basin (Batllori et al. 2015, 2019; Baudena et al. 2020) and boreal communities (Abis and Brovkin 2019), but none of them encompasses ecosystems throughout different biomes, with the notable exception of the seminal work of (Casagrandi and Rinaldi 1999).

In this work, we addressed the following research questions:

(RQ1) What set of characteristics can lead an individual PFT to persist, in isolation, for different emerging fire regimes?

(RQ2) Which are the main plant characteristics, if any, that influence the emergence of different communities?

(RQ3) What combination of plant characteristics can lead different plant communities to emerge as alternative ecological states?

# Methods

## *Model*

We developed a new conceptual model to describe the dynamics of fire-prone plant communities. Then, we numerically integrated the model equations, and we performed parameter sensitivity analyses to answer the three research questions listed above. This model is a generalization of the approach of Baudena et al. (2020), developed for Mediterranean forests.

We distinguished the PFTs by their main characteristics, focusing in particular on competitive ability (mostly representing shade tolerance), fire response (encompassing several traits from individual to PFT level) and vegetation flammability (here driving fire occurrence). These characteristics are represented and quantified by specific parameters. Here, the fire responses included both the resistance of individual plants during fire, e.g. due to a thick bark, and the post-fire recovery strategies at individual or population level, such as resprouting ability or the existence of a large, persistent and fire-resistant seedbank (Pausas and Keeley 2019; Miller et al. 2020).

In the model, each community was composed of three PFTs, which represent the most relevant plant types in a given ecosystem. We chose to limit the number of parameterized PFTs to three as a compromise between detail and parsimony, following the examples of e.g. Staver and Levin (2012), Abis and Brovkin (2019), and Batllori et al. (2015) for specific biomes. The model is space-implicit, i.e. it simulates the plant cover dynamics within an area (of the order of 100x100 m$^2$) where the seeds of the PFTs are assumed to be able to disperse homogeneously. Two factors drive the assembly dynamics: plant-plant competition (Sec. 2.1.1) and fire (Sec. 2.1.2). During fire-free periods, PFTs succession is regulated by plant competition for resources (mostly light in this work), following the approach of Levins (1969), Hastings (1980) and Tilman (1994). The (deterministic) succession is perturbed by fires, which are represented as stochastic events and occur in pulses. At each time, the chance of fire occurrence depends on the flammability of the community. In turn, different plant responses to fires lead to different post-fire community compositions. These two interactions create the fire-vegetation feedback in the model.

*Competition model*

Between two consecutive fires, the dynamics of the system is governed by three ordinary differential equations (Tilman 1994) for the variables $b_i$ ($i$=1,2,3), which represent the fraction of space occupied by PFT$_i$ ($0 \le b_i < 1$),

$$\frac{db_1}{dt} = c_1 b_1 (1 - b_1) - m_1 b_1 \tag{1}$$

$$\frac{db_2}{dt} = c_2 b_2 (1 - b_1 - b_2) - m_2 b_2 - c_1 b_1 b_2 \tag{2}$$

$$\frac{db_3}{dt} = c_3 b_3 (1 - b_1 - b_2 - b_3) - m_3 b_3 - c_1 b_1 b_3 - c_2 b_2 b_3 \, , \tag{3}$$

where $t$ represents time (in years, yr). Parameters $m_i$ are the plant mortality rates (yr$^{-1}$), while $c_i$ are the colonization rates (yr$^{-1}$), that represent the combined processes of seed production, germination, and establishment. Finally, $1 - \sum_i b_i$ is the amount of empty space. Each plant type can colonize both the empty spaces and the space occupied by the inferior competitors, where $c_i b_i$ is the fraction of space that PFT$_i$ can colonize per time unit. A fixed hierarchy between PFTs was assumed, from the strongest ($i = 1$) to the weakest ($i = 3$) competitor, corresponding to an inverse successional order (i.e., from late to early). The fractional cover of each PFT corresponds to the field cover in real ecological settings, which includes different layers, and is normalized to the total area of the layers. In the absence of fires, the plant community reaches a stationary state that can be easily determined (Tilman 1994).

*Fires*

Fires are modeled as instantaneous, stochastic events. These are represented by a nonstationary Poisson process: the average fire return time $T_f$ (yr) is exponentially distributed, and the process is "non stationary" because the average return time is state dependent (following e.g. D'Odorico et al. 2006), i.e. the value of $T_f$ changes across the simulation depending on the present community composition (see also Online Supplement C). As we consider a set climate in each ecosystem, the average fire return time is assumed to depend only on fuel availability and community composition, taking into account the different PFT flammabilities, as follows

$$T_f = \frac{1}{\sum_{i=1}^{3} b_i L_i}. \tag{4}$$

Hence, a larger plant flammability, $L_i$, determines more frequent fires. Similarly, abundant fuel (represented by large vegetation cover values, $b_i$) and in particular, a greater cover of the more flammable PFTs, decreases the average fire return time, thus leading to a higher chance of fires (D'Odorico et al. 2006; Baudena et al. 2010, 2020). The ecosystem is assumed to be fuel-limited, but not ignition-limited.

Since the exponential distribution of fire return times could lead to extremely frequent fires, we set the minimum fire return time, $T_f^{min}$, to 1 or 2 years, depending on the target ecosystem (see Table 1). This represents the time needed for a (partial) recovery of the ecosystem after fire, since burned ecosystems are not immediately prone to new fires. For numerical purposes, we also set the maximum fire return time to $T_f^{max} = 10^4$ yr.

At each fire event, the cover of each PFT$_i$ is instantly reduced, retaining only a fraction, $R_i$ (between 0 and 1), of the original cover before fire. The parameter $R_i$, called 'fire response' hereafter, accounts for different processes and plant strategies that can have complementary roles for PFT survival, including fire-related plant mortality and plant recovery strategies after fire. Following (Pausas and Lavorel 2003) we rated fire response strategies, assuming that strategies ensuring individual survival were more efficient (high $R_i$) than strategies resulting in PFT survival but individual loss after fire (intermediate $R_i$). For crown fires, which often completely burn the aboveground biomass, $R_i$ represents the efficiency of post-fire regrowth, due to resprouting or seedbank germination (Clarke et al. 2005, 2013). For surface fires, this parameter represents the persistence of plants during fire, e.g. thanks to a thick bark (Lawes et al. 2011; Pausas 2015b). In either case, the parameter $R_i$ rated the fire response of a PFT to the typical fire regime observed in the ecosystem where that PFT occurred. In this representation, the fire response parameter $R_i$ not only described the ability of plants to survive fires, but also implicitly included fire intensity, because it represented the severity of the fire and the strength of the response of a PFT to the typical fire activity of a certain geographical area. For simplicity, within a certain area, all fires were considered to have the same intensity, while across areas they could be different (e.g. typically crown fires in the Mediterranean and low-intensity surface fires for the savannas; Archibald et al. 2018).

Equation (4) and the fire response representation introduce a feedback between the probability of fire occurrence and the composition of the plant community: plant cover, which is affected by fires via $R_i$, in turn determines fire occurrence. As a consequence, we expect that in this model different fire histories occurring in an ecosystem may result in alternative ecological states, characterized by dissimilar communities (D'Odorico et al. 2006; Baudena et al. 2010; Kitzberger et al. 2012; Staver and Levin 2012).

We also defined a non-dimensional version of the model (see Online Supplement A for the explicit derivation), which allowed us to interpret the results obtained for RQ1. This non-dimensional model corresponds to a Lotka-Volterra's competition model, with null bottom-up competition, i.e. negligible effect of the weaker competitors on the stronger ones (Chesson 2000; Kot 2001; Rauschert and Shea 2017).

## Analyses

Before addressing the research questions, we performed a general analysis of the model dynamics, investigating the community composition and plant cover achieved in the absence of fires; we then activated the fire dynamics and assessed the long-term community structure and the possible presence of multiple, alternative ecological states.

To answer the research questions, the analyses included two parts: (i) 'PFT characteristics' (corresponding to RQ1), where we analyzed how the characteristics (i.e., the model parameters) of an individual PFT in isolation related to each other, and how these characteristics related to the resulting fire frequency in fire-prone environments; (ii) 'Community emergence' (corresponding to RQ2-3), where we assessed which characteristics, if any, of the PFTs present in a certain biome related to the emergence of different communities, possibly leading to alternative ecosystem states. To these ends, we explored the parameter space by running 50 simulations for each set of parameters, i.e. one point in the parameter space (see Sections 2.2.1 and 2.2.2.2), to capture the variability in cover due to the stochastic fire dynamics and initial vegetation cover. All simulations were run for 15,000 yr to ensure that the variability in vegetation cover generated by fire stochasticity was fully captured. We notice here that plant dynamics had much shorter time scales (see e.g. Fig. 1): in the 'Community emergence' analyses, the long-term ecological state was usually achieved in 100-1,000 yr for all the case studies considered; the convergence time was even shorter in the 'PFT characteristics' simulations.

### PFT characteristics

First, we studied the plant characteristics that can lead an individual PFT to persist in isolation, and the connections between these characteristics and the resulting fire frequencies (RQ1). To this end, we modeled the dynamics of a single PFT, by setting the cover of the other PFTs to zero. For these analyses we dropped the subscript $i$ for all the variables and parameters since only one PFT was considered in each simulation.

We generated random values of plant colonization rate, $c$, between 0.001 yr$^{-1}$ and 20 yr$^{-1}$, and for each of these we considered four values of the mortality rate, $m$, such that $c/m = [2, 5, 10, 20]$.

Then, for each combination of colonization and mortality rate, we varied the fire response, $R$, between 0.05 and 0.9 in steps of 0.05, and the flammability, $L$, between 0.001 yr$^{-1}$ and 0.99 yr$^{-1}$, increasing its value by 1.5 times at each step. Finally, for each parameter set, we run 50 different simulations by randomly varying the initial vegetation cover between 0.01 and 0.99.

We used the resulting average vegetation cover, $<b>$, as an indicator of the success of the PFT with the selected combination of $R, L, c$ and $m$. Since each fire event reduced the PFT cover, which instead grew between fires, we chose to compute the average vegetation cover of each simulation by using only the value right before each fire event in the last 20% of the total simulation time. For each parameter set, these values were then averaged across all the 50 runs. The same procedure was applied to compute the average fire return time, $<T>$, i.e. the average time between subsequent fires, representing the fire regime in our model. In Online Supplement A, we discuss how the non-dimensional version of the model helps interpreting the results.

## Community emergence

The second set of simulations was designed to assess the effect of plant characteristics in shaping plant communities (RQ2&3). We included three PFTs in this set of simulations, thus running the full model described in Sec. 2.1. To parameterize the model, we focused on three plant communities observed in different biomes where wildfires play a recognized role: Mediterranean forests and shrublands, tropical humid savannas and forests, and boreal forests. See Table 1 for a summary of the chosen plant types and their characteristics.

## PFTs and parameter settings

The hierarchy among the PFTs was established by considering juvenile and adult shade tolerance. The most competitive $PFT_1$ was usually a plant that can grow under scarce light availability. The $PFT_2$ could not survive at very low light levels but persisted more easily than the $PFT_3$ in partially shaded environments. The weakest competitor $PFT_3$ was affected by the shade of the other PFTs. The three PFTs in the three case studies were identified as follows.

We focused on the Mediterranean Basin as a representative example of the Mediterranean biome. We followed Baudena et al. (2020) in choosing Holm oak, *Quercus ilex,* as the most competitive, late successional $PFT_1$ (Acácio et al. 2007; Amici et al. 2013; Vayreda et al. 2016). The $PFT_2$ represented pine species, such as Aleppo pine, *Pinus halepensis*, and Brutia pine, *Pinus brutia* (Zavala et al. 2000; Zavala and Zea 2004), which are less shade tolerant than oaks. For the $PFT_3$ we chose a generic Mediterranean shrub seeder, simplifying from (Baudena et al. 2020) to represent a mix of *Rosmarinus*, *Cistus* or *Ulex* spp.

For the humid tropical regions, we simply captured the contrast between shade-tolerant, fire avoider rainforest trees ($PFT_1$) and shade-intolerant, fire resistant savannas, represented by savanna trees ($PFT_2$) and savanna C4 grasses ($PFT_3$) (Staver and Levin 2012; Charles-Dominique et al. 2018).

For boreal ecosystems, we focused on North American boreal species. We identified the shade tolerant balsam fir, *Abies Balsamea* (Uchytil 1991*a*) as $PFT_1$, and the less shade tolerant, but very common black spruce, *Picea mariana,* and jack pine, *Pinus Banskiana*, as $PFT_2$ (Carey 1993; Fryer 2014). These two latter species are similar in both shade tolerance and fire response. In the following we will refer to $PFT_1$ and $PFT_2$ as the fire avoider conifer and fire embracing conifer, respectively. The parameters estimated for balsam fir can also represent white spruce, *Picea glauca* (de Lafontaine and Payette 2010, 2012), which is a late successional, fire avoider conifer tree mostly found in western NA, where balsam fir is rare. Finally, shade intolerant deciduous broadleaved trees (Girardin et al. 2013) were chosen as $PFT_3$; specifically, we parameterized $PFT_3$ considering paper birch, *Betula papyrifera* (Uchytil 1991*b*), and trembling aspen, *Populus tremuloides* (Howard 1996).

The parameter values for each PFT (Table 2) were estimated as follows. Given the intrinsic ecological uncertainties in determining the parameter values, these were not intended to be exact values but rather as reference values, around which we performed the sensitivity analyses. See Online Supplement B for a detailed description of the PTF characteristics.

*Mortality rate.* We estimated mortality rates for each PFT as three times the inverse of the PFT average lifespan in the absence of competition and fires: if colonization is inhibited, plant cover decays exponentially, i.e. $b = b_0 \exp(-mt)$, thus reducing to $0.05 \, b_0$ within the average lifespan of the species.

*Colonization rate.* These parameters were defined following published estimates, together with additional information about growth time, spread rates and time needed to achieve a steady state after almost total plant burning (see Table S1 in Online Supplement B and references therein).

*Flammability.* We considered the typical fire return time in communities where the PFT represented the prevailing cover. When the domain is entirely covered by a certain PFT$_i$, eq. (4) gives $L_i = 1/T_f$ (since $\varepsilon$ is negligible), which defines the flammability as the inverse of the average fire return time in an ecosystem dominated by PFT$_i$.

*Fire response.* We classified plants into three main categories, having weak, intermediate and strong fire response (Pausas and Lavorel 2003; Jaureguiberry and Díaz 2023), which corresponded to different ranges of $R_i$. We used information on bark thickness, serotiny, post-fire resprouting strategies and rate of survival to frequent and intense fires to estimate this parameter. The range $0 < R_i < 0.3$ represented plants that do not have fire response strategies neither at plant nor at PFT level (i.e. fire avoiders); $0.3 < R_i < 0.7$ represented PFTs that do not display adaptations promoting individual adult survival, but have an extensive seed bank that survives fires allowing plant survival at PFT level (i.e. fire embracers); finally, $R_i > 0.7$ corresponded to plants having high individual fire resistance, due for example to resprouting ability or thick bark (i.e. fire resisters).

## Parameter sensitivity analyses

To answer RQ2&3, we performed a parameter sensitivity analysis, i.e. we explored the type of communities that emerged across the parameter space. The model includes twelve parameters (excluding the two small thresholds, $\varepsilon$ and $\delta$) and, among them, only the fire response ($R_i$) has a defined, limited range of variability. Hence, the corresponding parameter space is a potentially infinite hypervolume. We limited the parameter space to exclude unrealistic parameter combinations and explored it around the reference values identified for the real communities in the three case studies described in Section 2.2. Fire responses, $R_i$, were varied in the range 0.01-0.90 in steps of 0.02, while $c_i$, $m_i$ and $L_i$, were varied in a realistic broad range, from 0.5 to 2 times the reference value (flammability was increased by 1.05 its value at each step, while colonization and mortality rates were varied dividing the explored range into 40 steps). Colonization rates $c_i$ were always chosen to be larger than the mortality rates $m_i$ to ensure plant survival in isolation without fire (Tilman, 1994).

For each parameter set, we ran 50 simulations with different initial vegetation cover of the PFTs included in the community. We ensured that the total initial vegetation cover of the three PFTs was $\sum b_i \leq 1$ (Tilman 1994), by generating three random numbers in the range between 0.01 and 0.99, and then dividing each of them by their sum. The values thus obtained were arbitrarily assigned to the three PFTs. These runs allowed us to: i) account for the variability due to fire stochasticity; ii) observe all of the possible communities that could be achieved for a certain set of parameters owing to the fire-vegetation feedback (i.e. the fire return time changed as a function of the vegetation cover and community composition, possibly resulting in different trajectories, and leading to alternative ecological states). Hence, the community can be reset by fire and change thereafter owing to the plant succession. In each run, we recorded the community composition before every fire event in the last 20% of the simulation time (15,000 y), considering only the PFTs that had $b_i \geq 0.03$. This procedure was especially relevant in case of recurrent alternance between different communities along the time series. We thus obtained a compilation of the possible states (i.e., communities) achieved in the time series, for each parameter set.

To answer RQ2, we varied the parameters of each PFT, changing one parameter value at a time. In the explored range, the parameters associated with the largest community changes with respect to the reference were interpreted to be the plant characteristics that were most relevant for determining the system state (i.e. the community composition).

Concerning RQ3, we explored parameter-space sections obtained by varying selected couples of parameters among the most relevant ones (in the sense of the analysis described above), while keeping all other parameters at their reference value. This allowed us to identify the parameters that were most relevant for determining the existence of multiple alternative ecological states.

Finally, in a subset of the simulations, we accounted for the possible arrival of seeds from surrounding areas (e.g., due to wind or animal transport), preventing a certain PFT from disappearing after fire. We thus set a minimum post-fire vegetation cover $\delta \simeq 10^{-4}$, representing germination of seeds coming from outside the study area.

# Results

## *PFT characteristics*

We first examined the dynamics of a single PFT. In this set of simulations and in the explored parameter range, the system displayed only one final state (i.e., multi-stability, corresponding to either a vegetated or not vegetated state, was never observed). When fires substantially reduced the plant cover (e.g., at low fire response values), the average fire return time rose (Figure 1A-B) owing to the fire-vegetation feedback, which allowed the PFT to re-establish itself. Therefore, we never observed the total die out of the PFT.

To answer RQ1, we used two illustrative examples of PTFs with either high (0.3yr $^{-1}$, Fig 1a,c) or low (0.05yr $^{-1}$, Fig. 1b,d, right panels) colonization rate *c* (mortality rate was *m*=0.1*c*); similar results were obtained for other combinations of colonization and mortality rates (as described in Sec. 2.2.1). When the plant flammability *L* was low, fires were rare (*<T>* of 100 yr or larger, Figure 1A-B, light areas), and the average PFT cover was similar to its equilibrium value in the absence of fires (*<b>* close to 0.9 in both examples, Figure 1C-D, dark areas) for any fire response *R*. On the other hand, when the flammability was high, both the fire interval (upper half of Fig. 1A-B), and the average cover (upper half of Fig. 1C-D) depended on the specific fire response of the plant. The average cover of fire intolerant (low *R*) and highly flammable PFTs was strongly reduced by the resulting regime of frequent fires (top-left corners of Fig. 1A-B and Fig. 1C-D). In contrast, frequent fires only slightly reduced the cover of fire resistant PFTs (high R). In other words, plants having a strong fire response could effectively maintain a high cover for any flammability and fire frequency; while plants having a weak fire response displayed a lower flammability that led to infrequent fires, thus allowing for the plant spread and resulting in high plant cover values. Furthermore, the response-flammability relationship was stronger for slow than for fast colonizer plants: given a *(R, L)* pair value, the average vegetation cover depended on the time scale of plant colonization. Faster colonizing PFTs had a greater cover (Fig. 1C) than slower ones (Figure 1D), despite the higher fire frequency associated with the former (Figure 1A) compared to the latter (Fig. 1B).

The effect of the plant colonization parameter (related to the time scale of plant growth) can be explained by using the non-dimensional formulation of our model (Eq. S2-S4 in Online Supplement A). If the rescaling of the non-dimensional model is applied to Fig. 1, i.e. using the non-dimensional flammability $L/c$ as a vertical axis and representing the rescaled vegetation cover $b\,c/(c-m)$, then Fig. 1C becomes the same as Fig. 1D. This is shown in Fig. S1 (Online Supplement A), for a set of non-dimensional flammability values and different combinations of $R$, $L$, $c$ and $m$.

## *Community emergence*

In the absence of fire, a closed canopy forest of the late successional $PFT_1$ tree established in all the three case studies, i.e. for the reference parameter values reported in Table 2. When including fires, only the Mediterranean case study preserved this stable state, while the tropical and boreal cases showed alternative ecological states (Figure 2-4). In particular, in the Mediterranean community, the evergreen, fire-resistant oak ($PFT_1$) eventually outcompeted the other PFTs by maintaining a long fire return time (Fig. S3A in the Online Supplement) that averaged to 490 yr (in line with Baudena et al. 2020; Vasques et al. 2022). For the tropical communities, depending on the initial condition and on the specific stochastic fire sequence, we observed the establishment and maintenance of either a closed canopy forest ($PFT_1$) or a mix of savanna trees and grasses ($PFT_2$+$PFT_3$). These alternative states respectively corresponded to average fire return times of 1045 yr and 4 yr (Fig. S4A in the Online Supplement), in line with observations (e.g., D'Onofrio et al. 2018). Such a result was expected as humid savannas and tropical forests are observed in areas with the same environmental conditions but different fire frequencies (Hirota et al. 2011; Staver et al. 2011*b*; Dantas et al. 2016), which is commonly explained as an indication of alternative biome states maintained by a fire-vegetation feedback (e.g., Accatino et al. 2010; Staver and Levin 2012). In the boreal case study, fires triggered the irregular alternation between forests of either embracer conifers ($PFT_2$), avoider conifers ($PFT_1$), or a mixedwood forest including deciduous trees and late successional evergreen conifers ($PFT_1+PFT_3$). This alternation persisted through the whole time series, see Figure 2, creating recurrent but irregular sequences of states. Mosaics of different plant communities are commonly observed in boreal North America landscapes (Bormann and Likens 1979; Baker 1989; Gumming et al. 1996; Johnson et al. 1998; Weir et al. 2000). According to our simulations, the mosaic would be generated by the combination of: i) the irregular alternation between communities sustained by different fire frequencies within each patch (more frequent fires in embracer conifer forests, with 80 yr average fire return time, and less frequent fires in mixedwood or late successional conifer forests, with respectively about 100 and 450 yr average fire return time), and ii) asynchrony in fire dynamics (matching differing and independent fire histories) between patches of the same landscape. This asynchrony is similar to the gap dynamics observed by Wissel (1992). In addition, the communities reported here are in line with field observations of most common communities (Jasinski and Payette 2005; Couillard et al. 2012) and paleoecological findings, which identify recurrent turnovers between balsam fir and black spruce communities, characterized by fire frequency shifts (Ali et al. 2008; de Lafontaine and Payette 2010; Couillard et al. 2018).

*Main characteristics shaping plant communities*

The parameter sensitivity analysis (RQ2) showed that a few key plant characteristics, depending on the biome, influenced the emergence of different communities in this model (Fig. 3 and Sec. 2.3.2). Such emergence depended on the ability of a PFT to persist given the constraints imposed by the other plants, in terms of competition for resources and fire frequency.

In the Mediterranean case study (left panels in Fig.3), a state change was observed only for low fire response values of the strongest competitor $PFT_1$ ($R_1$), where the $PFT_1$ forest became bistable with a shrubland, $PFT_3$. Conversely, for medium-to-high values of $R_1$ or for changes in all other parameters, the $PFT_1$ forest was the only observed state.

In the tropical (central panels in Fig. 3) and boreal (right panels of Fig. 3) case studies, the picture was more complex. State changes were observed for variations in the colonization rate of nearly all PFTs, except for $PFT_3$ in the boreal case. On the other hand, modifying the mortality rates led to fewer state changes (i.e., $m_2$ in the tropical case and both $m_1$ and $m_2$ in the boreal case). For each of the PFTs, a $c_i$-$m_i$ relationship emerged in the model: the ecological states observed when increasing the colonization rates with respect to the reference values (black vertical lines in Fig. 3) were similar to the ones observed when decreasing the mortality rate, and vice versa (see Fig. S6 in the Online Supplement). Yet, the colonization rate led to the highest number of overall state changes. No changes were observed in the explored $L_1$ and $L_2$ ranges, while only large values of $L_3$ in the tropical communities led to state changes. Concerning the fire responses, a $PFT_1$ forest was always present at large $R_1$, while bistable states ($PFT_1$/$PFT_2$+$PFT_3$ for the tropical community) or temporal alternation between states ($PFT_1$/$PFT_2$/$PFT_1$+$PFT_3$ for the boreal community) appeared at low $R_1$, in analogy with the Mediterranean case. State changes emerged at low values of $R_2$ for the tropical case only, while no state change was observed in the whole $R_3$ range.

In conclusion, the parameters that mostly changed the long-term ecological state across the three communities were: $R_1$, $c_1$ and $c_2$.

## *Combinations of plant characteristics leading to alternative ecological states*

To answer RQ3, we explored three plane sections in parameter space, each defined by pairs of the three parameters selected in Sec 3.2.1 (i.e. $R_1$, $c_1$ and $c_2$). Among those, we observed few state changes when expanding around the reference values of the $PFT_1$ and $PFT_2$ colonization rates, i.e. the $c_1$-$c_2$ plane shown in Fig S7 in the Online Supplement. As expected from the earlier analysis (Fig. 3), a $PFT_1$ forest was the only state observed in the whole $c_1$-$c_2$ plane (Fig. S7A) for the Mediterranean case. In the tropical (Fig. S7B) and boreal (Fig. S7C) cases, stable states emerged at either high or low values of $PFT_1$ colonization rate, while bistable states or irregular alternance of states respectively emerged for the two cases at intermediate values of $c_1$. Since the Mediterranean case suggested a decisive effect of the fire response of the strongest competitor, $R_1$, on the stability of the observed communities, we further focused the analysis on the $R_1$-$c_2$ and $R_1$-$c_1$ planes (Fig. 4).

*The fire response of PFT₁ ($R_1$) and the colonization rate of PFT₂ ($c_2$)*

In the Mediterranean biome, when exploring the $R_1$-$c_2$ plane (Fig. 4A), the dominance of PFT$_1$ was maintained at large R$_1$ values for any value of $c_2$. At low $R_1$ values, the PFT$_1$ forest became bistable with other states, whose specific composition depended on the PFT$_2$ colonization rate. There, the weak fire response of PFT$_1$ and its slow colonization time scale compared to the other PFTs made it susceptible to die out under high fire frequency. Across the time series, if the cover of the other, more flammable, plants became large enough to maintain a short fire return time, PFT$_1$ would succumb. If, otherwise, PFT$_1$ dominated the community, a low fire frequency was maintained, allowing this late successional plant to outcompete the others during the fire-free periods. When bistability was possible, we observed that whether the system ended up in one or the other state depended on the initial plant cover of the community and on the specific sequence of stochastic fires that might lead to a series of short or long fire intervals (see also Online Supplement C). Similar dynamics drove the occurrence of alternative ecological states also in the other case studies. Tri-stability was observed at the borders between areas of different types of bistability.

The bistability between tropical forest and savanna was observed in a broad part of the $R_1$-$c_2$ plane (Fig. 4C), where the fire response of PFT$_1$ was low. Remarkably, the pattern of states in the tropical parameter space was not dissimilar from the Mediterranean case (compare Fig. 4C with Fig. 4A): a broad area of PFT$_1$ dominance was observed at large $R_1$, bistable states were possible at intermediate to low $R_1$, and regions of tri-stability occurred at the edges between these areas of bistability. This similarity of behavior was interesting, as the characteristics of the Mediterranean PFTs were substantially different from the corresponding tropical PFTs. The latter displayed faster dynamics (given by $c_i$ and $m_i$), stronger fire response of PFT$_2$, and higher flammability of PFT$_2$ and PFT$_3$.

In the boreal biome, the irregular alternation of forests of either late successional conifers (PFT$_1$), embracer conifers (PFT$_2$), or a mixedwood forest (PFT$_1$+PFT$_3$) was observed in a narrow region of the parameter plane, at low $R_1$ values. In analogy with the other case studies, the temporal alternation of different communities was observed at intermediate to low $R_1$, while stable ecological states were observed at high $R_1$ values (Fig. 4E). These stable states corresponded to either a $PFT_1$ forest, at low colonization rate of $PFT_2$, or to a $PFT_1+PFT_2$ forest, at high colonization rate of $PFT_2$. The existence of two stable states at large $R_l$ differed from the other case studies, and it follows from Tilman's (1994) model. In the absence of fire, $PFT_2$ can coexist with $PFT_1$ when its colonization rate is fast enough to compensate for losses due to competition pressure, i.e. $c_2 > c_1(c_1 - m_1 + m_2)/m_1$ . Using the reference values for $c_l$, $m_l$ and $m_2$ (Tab. 2), this condition corresponds to $c_2 > 0.158$ yr$^{-1}$ in the boreal case, which closely corresponds to the value of $c_2$ that separates the $PFT_1$ and $PFT_1+PFT_2$ states at large $R_l$ in Fig. 4E.

For all three case studies, the ecological states observed at large $R_l$ values were the same as those predicted by Tilman's unperturbed model for the $c_i$ and $m_i$ values used in that region of parameter space. This corresponded to a stable $PFT_1$ state for both the Mediterranean and tropical cases, where Tilman's coexistence condition is never satisfied in the explored range of $c_2$ parameter values. In contrast, Tilman's coexistence condition is met in part of the explored range of Fig. 4E, and both $PFT_1$ and $PFT_1+PFT_2$ states are observed at large $R_l$ values. Consequently, the competition between PFTs mostly drove the community dynamics at large $R_l$ values.

*The fire response of $PFT_1$ ($R_l$) and the colonization rate of $PFT_1$ ($c_l$)*

We observed a relationship between the colonization rate and the fire response of $PFT_1$ (Fig 4B, D, F). As $c_l$ values increased, the stable state region characterized by $PFT_1$ alone (in the Mediterranean, boreal and tropical communities) or $PFT_1$ coexisting with $PFT_2$ (in the boreal community) became wider, progressively extending towards low $R_l$ values. Hence, in order to maintain a stable fire community, $PFT_1$ needs to be either very fire resistant (large $R_1$) or very fast in its colonization rate (large $c_1$), as this also ensures rapid expansion after fire. Conversely, a fire intolerant $PFT_1$ could be lost from the community if its colonization ability was insufficient, despite its superior competitive ability. This was, for instance, observed at low $R_l$ and low $c_l$ values in the Mediterranean and boreal communities, where respectively stable states of $PFT_3$ and $PFT_2$ emerged. In the boreal case (Fig. 4F), we also observed a sharp transition at about $c_l$=0.07yr$^{-1}$, that identified a state change between $PFT_1$ and $PFT_1+PFT_2$ forests at large $R_l$. This again matched Tilman's coexistence condition (Tilman, 1994) for the no-fire dynamics, as recalled above, i.e. $c_l < 0.078$ yr$^{-1}$ using the reference values of $c_2$, $m_2$ and $m_l$ (Tab. 2).

The relationship between the $PFT_1$ colonization rate and its fire response was confirmed for concomitant variations of the $PFT_1$ and $PFT_2$ colonization rates. In a version of the $R_1$-$c_2$ parameter plane simulated for a lower value of $c_1$, the colonization rate of the strongest competitor $PFT_1$ (Figure S8 in the Online Supplement), the patterns of states described above (Figure 4 A,C,E) seemingly shifted towards higher values of $PFT_1$ fire response for all the biomes (thus towards the right in the figures). At low fire response values, new (mostly stable) ecological states emerged.

Finally, the inclusion of seed spreading from the surrounding environment originated irregular alternation among the ecological states in the Mediterranean and tropical cases. However, this hardly affected the ecological patterns shown in Fig. 4. In particular, the communities involved in the multiple ecological states of Fig. 4 could alternate across the time series (similar to that observed in the boreal case and in Casagrandi and Rinaldi 1999), while regions of parameter space where only one ecological state was possible were not affected by the inclusion of the seed inflow from neighboring areas.

# Discussion

In the model presented here, the fire response emerged as a key plant characteristic influencing PFT density, fire frequency and the existence of one or more communities in different fire biomes. For single PFTs (RQ1), less flammable plants can have a high cover over a wide range of fire responses, while highly flammable plants have to display a strong fire response to maintain a high cover. In a competitive fire-prone community, the fire response of the strongest competitor determined whether only one or multiple alternative ecological states were feasible (RQ3): a strong fire response resulted in the existence of only one stable state, whereas a weak fire response allowed for the existence of alternative ecological states. In the latter case, the colonization rate of both the strongest competitor and the second-best competitor explained the observed communities in different fire prone ecosystems (RQ2).

In the absence of competition from other plants, the vegetation cover and the fire frequency of a single standing PFTs shaped each other via the fire-vegetation feedback. A relationship between the fire response of a single PFT and its flammability spontaneously emerged in our model, and this relationship was mediated by the PFT colonization time scale (similar to Jaureguiberry and Díaz 2023). In detail, less flammable plants generally led to rare fires and maintained a large cover over a broad range of fire responses. Conversely, highly flammable plants, leading to frequent fires, can maintain a high cover if they are slow growers and fire tolerant (high values of parameter $R$), while they can display an intermediate to strong fire response if they are fast colonizers. The plant colonization rate, together with the fire response, determined the average fire frequency and plant cover: PFTs that spread rapidly were also fast at recovering after fires, even when the fire response of the plants was weak, thus ensuring high plant cover (Eq. 4).

Real plant behavior supports the model results. According to ecological observations, resprouting species are indeed present in ecosystems having various fire frequencies, spanning from the flammable eucalyptus forests in Australia (STRASSER et al. 1996) to the less flammable oak forests in the Mediterranean basin (see e.g., references in Table 2); although, resprouting attributes are more common in ecosystems where the fire frequency is higher, especially for woody plants (Harrison et al. 2021). Conversely, fire intolerant trees, such as tropical forest trees, generally create a moist understory and therefore decrease ecosystem flammability. Once infrequent fires allow the establishment of those trees, they maintain a low ecosystem flammability that creates a positive feedback, further allowing their spread. On the other hand, in fire-prone environments, fast colonizers, such as grasses or early successional shrubs, are often highly flammable, although their fire response can span from the highly fire-adapted savanna and Mediterranean grasses (see e.g., Baudena et al. 2020) that can resprout, to the weak fire response of some annual grasses, such as cheatgrass (*Bromum tectorum*; Zouhar 2003). This latter grass is infamous for its success in invading North American prairies and changing their fire regime (Fusco et al. 2019), yet it does not resprout and has seeds that are susceptible to heat kill (but can survive underground when already present). The fast colonization rate of field layer species guarantees their survival independently of their fire response, as long as some individuals and propagules are not totally burned. Our findings indicate that these types of behavior might be due to fires acting as a filter on species characteristics and their possible associations. This agrees with evolutionary theories and modeling outcomes suggesting that flammability, despite its negative effects on individuals, may have evolved in combination with other advantages, such as higher recruitment opportunities or resprouting (Bond and Midgley 1995; Schwilk and Kerr 2002). It should be noted that, despite the similarities between the trait syndromes found in this work and in evolutionary studies, in our model we cannot establish a causal relationship between these characteristics.

Within the framework of the model, plant survival in a community was further constrained by the features of other species adapted to the same environment. In particular, the characteristics of the most competitive PFT were found to be of primary importance for community composition in the long term (Figure 2). When the strongest competitor also had a strong fire response, only one type of community was possible: a forest of this late successional tree (in the Mediterranean and tropical case study), which could coexist with the embracer conifer in the boreal case study. Conversely, when the strongest competitor had weak to intermediate fire response, two or more alternative ecological states were possible (Figure 2), and the specific composition of the observed communities mostly depended on the colonization rates of the two best competitors. Here, we also note that the colonization rates of the two best competitors (together with their mortality rates) determined the competition strength of the first on the latter (i.e., of $PFT_1$ on $PFT_2$), as shown by the derivation of the non-dimensional model (Eq. S6 in Online Supplement A). The combination of the fire response of the strongest competitor and the competition strength between the two best competitors explained the presence of a stable forest of resprouter holm oak in the Mediterranean basin (Amici et al. 2013; Carnicer et al. 2014), the bistability between the fire-intolerant tropical forest and humid savannas (Staver et al. 2011$a$; Dantas et al. 2016; D'Onofrio et al. 2018) and the temporal alternation of fir- or spruce-dominated forests reported for North America by palaeoecological records (Couillard et al. 2018). Moreover, the drivers that we identified for the plant communities agree with, and expand on those used by Van Nes et al. (2018), which explain the forest-savanna bistability as a tradeoff between growth and fire-induced mortality of trees.

The factors that mostly decide the community composition in this model closely correspond to the classification of plant persistence conditions proposed by Pausas and Lavorel (2003). This framework suggests that, in a fire prone and competitive community, plant attributes determine the possibility of plant persistence at either individual, population, community or landscape level.

While our model did not have explicit representation of these persistence levels, and therefore $PFT_i$ persistence corresponds only to a positive fractional cover ($b_i \neq 0$), we can connect persistence levels to model parameters. Persistence of individuals after fire (e.g., due to resprouting) was represented by large values of the fire response parameter, while population persistence (e.g. due to a fire-resistant seed-bank) corresponded to intermediate-to-low fire response values. In the model, the fire response of the strongest competitor was also a key factor in determining the ecological states. The persistence condition at community level, i.e. species survival in a competitive environment between fire events, is due to competitive ability in Pausas and Lavorel's framework. Analogously, in our model, the competition strength between the two best competitors (determined by the colonization and mortality rates of $PFT_1$ and $PFT_2$, see Eq. S6 in Online Supplement A) is crucial for determining the community composition. Finally, the framework predicted that the explicit representation of external seed dispersal might lead to species persistence at landscape level. In our model, seed dispersal could lead to temporal alternation between states, but had the weakest effect on the community composition. This correspondence with the early conceptual work of Pausas and Lavorel (2003) brings ecological support to our results. Furthermore, since the importance of these persistence levels emerged from the dynamics of the model rather than being strictly imposed, our results provide quantitative support to such a theoretical ecological framework. In addition, we showed that different persistence strategies could be related, such as the individual/population ($R_i$) and community ($c_i$) persistence levels (Figure 4B, D, F): the competition strength depended on the colonization rate of the PFTs, which in turn regulated the regrowth time between consecutive fires, thus having a complementary role with the fire response of a certain PFT.

Sequences of long (or short) fire return times, which randomly occurred in the fire series, often started the exclusion of PFTs and the transition between different communities (see Online Supplement C). These sequences triggered the fire-vegetation feedback, which resulted in the state change. For stable or bistable communities, as in the Mediterranean and Tropical cases, such sequences played an important role in the initial part of the simulations, when they allowed a community to become dominant. At that point, the established community was maintained in time by the fire-vegetation feedback, while the PFTs not included in the community eventually died out. This agreed with observations suggesting that accidentally frequent (or infrequent) fires can prevent (or foster) transitions between stages of tropical forest development (Lehmann et al. 2011; Hoffmann et al. 2012; de L. Dantas et al. 2013). In the case of irregular alternances, the transitions between states occurred across the whole simulation time, as can be seen in Fig. 2 (e.g. between 10 kyr and 13 kyr). In this case, all the PFT covers were positive along the whole simulation time, despite sometimes reaching very low values ($b \approx 0.001$). Hence, specific sequences of short or long fire return time allowed the blowing up of one or more of the PFTs and caused a (temporary) state change. The same mechanism triggered alternation between states when including seed dispersal. Because an increasing frequency of extreme events is expected according to climate change projections (Keeley and Syphard 2019; Masson-Delmotte et al. 2021), it becomes important to include stochastic effects for studies on possible state transitions in fire-prone communities.

Although the model does not explicitly include climate factors, regional climate gradients or climate change are expected to modify plant behavior, as represented by the model parameter values. Hence, we can use the results of the sensitivity analysis to speculate on the possible changes of fire-prone ecosystems in response to climate variations. For instance, black spruce ($PFT_2$) dominates in the western part of North America, where the climate is drier than in the eastern area, resulting in the slowing down of balsam fir ($PFT_1$) establishment and growth (Frank 1990; Goldblum and Rigg 2005). This would correspond to the reduction of the $PFT_1$ colonization parameter ($c_1$) in the model, i.e. downward migration of the black cross in Fig. 4F, which explains the loss of $PFT_1$ (balsam fir) and the dominance of $PFT_2$ (black spruce) observed in the western boreal North American regions. Similar exercises may give indications on possible community changes under climate change scenarios. Global warming, changes in precipitation regimes and a general tendency towards increasing aridity and drought occurrence are predicted over large portions of the planet (Masson-Delmotte et al. 2021), which will affect plants, fires and their interactions (Bradstock 2010; Higgins and Scheiter 2012). As a rule of thumb, the predicted changes are expected to slow down plant colonization rates, increase plant flammability and possibly reduce plant fire response, thus shifting the modeled systems towards the bottom left corner in Fig. 4. This would for instance lead the Mediterranean oak forest to become bistable with an open shrubland (Fig 4A-B) and even disappear completely under the most extreme reductions (Fig. 4B and Fig. S8A in the Online Supplement), in line with previous findings (Batllori et al. 2019; Baudena et al. 2020).

This model is expected to be adaptable to any fire ecosystem worldwide. Clearly, it is also a simplified representation of real ecosystems. For example, fire response may not be constant in time: some seeders, such as pines in the Mediterranean Basin (Pausas 1999*a*; Climent et al. 2008), only produce seeds when mature, resulting in a demographic bottleneck if a second fire occurs before maturity is reached. Likewise, no intraspecific dynamics are accounted for, and spatial processes are not represented, while spatial vegetation patterns might prevent the occurrence of alternative ecosystem states (Rietkerk et al. 2021). Despite these simplifications, our results agree with the findings of models that explicitly represent seedbank dynamics (Baudena et al. 2020) (or spatial processes (Vasques et al. 2022). In addition, the limited number of parameters make this model an efficient conceptual framework, which can also be examined analytically in some cases (Baudena et al. 2020).

Despite the variety of models accounting for fire dynamics (Williams and Abatzoglou 2016), the degree of complexity that is required to capture the main features of fire-prone ecosystems is still unclear. In this context, our study underlines the importance of representing plant fire response. An improved representation of plant post-fire recovery led to a better reproduction of the forest burned area observed in western US (Abatzoglou et al. 2021). The representation of plant fire strategies improved simulations of fire regimes in Australian savannas using LPX (Kelley et al. 2014), a Dynamic Global Vegetation Models (DGVMs, e.g., Prentice et al. 2007). More generally, DGVMs often account for only a hurried conceptualization of post-fire recovery, and do not include resprouting as a trait (Kelley et al. 2014; Hantson et al. 2016; Venevsky et al. 2019; Harrison et al. 2021). We envisage that an improved representation of fire response could reduce projection uncertainties and assist also in ecosystem management and landscape planning for fire prevention (Hantson et al. 2016).

# Acknowledgements

We thank Mariska te Beest, Sandy Harrison, Ángeles G. Mayor, Max Rietkerk and Maria J. Santos for critical comments on a proposal, which partly led to this manuscript. We thank Sally Archibald for comments on an early version of the manuscript. RDS acknowledges support from the Spanish Government, grant number 453 RTI2018-094691-B-C31 (MCIU/AEI/FEDER, EU). MM acknowledges support by the Swaantje Mondt PhD Travel Fund of the Centre for Complex System Studies (CCSS) of the Utrecht University to visit MB. LS acknowledges support from the Leverhulme Centre for Wildfires, Environment and Society. AP acknowledges support from the EU H2020 project FirEUrisk, Grant Agreement Number 101003890. MB and AP acknowledge the Italian National Biodiversity Future Center (NBFC): National Recovery and Resilience Plan (NRRP), Mission 4 Component 2 Investment 1.4 of the Ministry of University and Research; funded by the European Union – NextGenerationEU (Project code CN00000033). The CCSS is gratefully acknowledged also for computing resources of the Clue facility.

# Statement of Authorship

MM contributed to the software development, coding simulation, model parameterization, analysis of results, visualization, writing the original draft, reviewing, and editing. RDS derived the nondimensional formulation of the model, and contributed to the analysis of results, writing the original draft, reviewing, and editing. LS contributed to conceiving the original idea of this work, writing the original draft, the reviewing and editing. AP provided useful feedback during result analysis, and contributed to supervision, reviewing, and editing. MB conceived the original idea of this work, and contributed to software development, model parameterization, analysis of results, supervision, visualization, writing the original draft, reviewing, and editing.

# Data and Code Availability

The simulation code and relevant data for the reproduction of results presented here are freely available on Zenodo repository https://doi.org/10.5281/zenodo.7763275.

# References Cited Only in the Online Enhancements

# Tables

**Table 1.** Characteristics of each PFT in the three case studies: Mediterranean, Tropical and Boreal North America biomes.

| Biomes and location | Mediterranean forest and scrubs (Mediterranean basin) | | |
|---|---|---|---|
| **PFT identification** | **PFT1**: *Quercus ilex* | **PFT2**: *Pinus* spp. | **PFT3**: *Rosmarinus officinalis, Cistus* spp.*, Ulex* spp |
| **Shade tolerance** | **High**: late-successional, broad-leaved evergreen | **Medium**: outcompeted by *Quercus ilex* (in mesic conditions) | **Low**: shrubs and perennial grasses outcompeted by trees |
| **Fire response** | **Strong**: high individual fire resistance, resprouts readily after fires | **Intermediate**: resistance at population level, post-fire seed release, serotinous cones in some species, thin juvenile bark, cones destroyed by severe crown fires, seedbank not produced in case of frequent fires preventing pines to reach maturity | **Intermediate**: shrubs: seeders with large seed banks and gradual seedbank decay, affected by ground fires; *B. retusum*: resprouter grass |
| **Flammability** | **Low**: moist and cool understory, very little fine and dead standing fuel, infrequent fires, average fire return time of 500 yr | **Intermediate**: vertical continuum fuel, flammable litter accumulation on the ground owing to summer senescence of needles, average fire return time of ~50yr | **High**: average fire return time of ~10yr |
| **References** | (Acácio et al. 2007; Puerta-Pinero et al. 2007; Saura-Mas et al. 2009; Tinner et al. 2009; Baeza et al. 2011; Amici et al. 2013; Azevedo et al. 2013; Clarke et al. 2013; Zeppel et al. 2015; Vayreda et al. 2016; Baudena et al. 2020) | (Pausas 1999a, 1999b; Zavala et al. 2000; Zavala and Zea 2004; Rodrigo et al. 2007; Climent et al. 2008; Fernandes et al. 2008; Karavani et al. 2018) | (Hanes 1971; Baeza et al. 2006; Acácio et al. 2009; Santana et al. 2010; Acácio and Holmgren 2014) |

| Biomes and location | Tropical savanna and forest | | | Boreal forest (North America) | | |
|---|---|---|---|---|---|---|
| | **PFT1**: Forest trees | **PFT1**: Savanna trees | **PFT3**: C4 grasses | **PFT1**: *Abies balsamea, Picea glauca* | **PFT1**: *Picea mariana, Pinus Banskiana* | **PFT3**: *Betula papyrifera, Populus tremuloides* |
| **PFT identification** | | | | | | |
| **Shade tolerance** | **High**: broad leaved | **Medium-Low:** juvenile light sensitivity | **Low**: grows in canopy gaps or coexists with trees having low brunch density (e.g. savanna trees) | **High**: late successional trees | **Intermediate:** replaced by late successional species | **Low:** pioneer species, rare in late successional communities, can grow in forest canopy gaps |
| **Fire response** | **Weak:** low resistance at PFT level, fire-sensitive trunk | **Strong:** resistance at individual level, thick bark, survive to frequent ground fires | **Strong:** resistance at individual level, strong resprouters | **Weak:** low resistance at PFT level, avoider, seeder, seeds often destroyed by fires, survival trough unburned adult trees | **Intermediate:** high resistance at PFT level, embracer, semi-serotinous cones | **Strong:** high resistance at individual level, resister, resprouts form root collar when killed by fires |
| **Flammability** | **Low:** humid and shady understory, infrequent fires, average fire return time of ~1000yr | **High:** frequent fires, average fire return time of 3-5yr | **High:** fast-spreading ground fires occurring possibly every year | **Low:** fires almost absent in *Abies b.* forests, average fire return time of 250-300yr | **Intermediate:** layered structure, resinous wood, average fire return time of ~75yr | **Low:** high canopy moisture content, infrequent fires, average fire return time ~100yr |
| **References** | (San José and Farinas 1991; Kurokawa et al. 2003; Laurance et al. 2004; Russell-Smith et al. 2004; Bampfylde et al. 2005; Rossatto et al. 2009; Geiger et al. 2011; Viani et al. 2011; Hoffmann et al. 2012; de L. Dantas et al. 2013; D' Onofrio et al. 2018) | (Beckage et al. 2009; Warman and Moles 2009; Accatino et al. 2010; Baudena et al. 2010; Lehmann et al. 2011; Hoffmann et al. 2012; de L. Dantas et al. 2013) | (Accatino et al. 2010; Baudena et al. 2010) | (Rowe and Scotter 1973; Uchytil 1991a; Galipeau Carey 1993; Larsen 1997; Sirois 1997; Greene and Johnson 2000; Bergeron 2000; Ali et al. 2008; Bouchard et al. 2008; de Lafontaine and Payette 2010, 2012; Couillard et al. 2012; Abrahamson 2014; Rogers et al. 2015) | (Viereck 1983; Payette 1992; Payette 1992; Larsen 1997; Sirois 1997; Greene and Johnson 1999; Laberge et al. 2000; Amiro et al. 2001; Wirth 2005; Ali et al. 2008; Bouchard et al. 2008; de Lafontaine and Payette 2010; Johnstone and Payette 2010; Payette et al. 2012; Fryer 2014; Héon et al. 2014; Rogers et al. 2015; Couillard et al. 2018; Hart et al. 2019, 2019) | (Swain 1980; Uchytil 1991b; Zasada 1992; Payette 1993; Howard 1996; Larsen 1997; Greene and Johnson 1999; Bergeron 2000; Brassard and Chen 2006; Bouchard et al. 2008; Bergeron and Fenton 2008; Rogers et al. 2013; Hart et al. 2019; Mack et al. 2021) |

**Table 2.** Reference parameter values of colonization rate ($c_i$), mortality rate ($m_i$), flammability ($L_i$) and fire response ($R_i$) of $PFT_{1,2,3}$ as parameterized for Mediterranean, tropical and boreal communities. Possible ranges identified in the parameterization (Online Supplement B) are reported in parentheses.

| Parameter | Mediterranean | Tropics | Boreal | Units |
|---|---|---|---|---|
| $c_1$ | 0.047 | 0.20 (>0.15-2.50) | 0.085 | |
| $c_2$ | 0.053 | 0.15 (0.15-2.50) | 0.13 | $yr^{-1}$ |
| $c_3$ | 0.3 | 20 (20-200) | 0.17 | |
| $m_1$ | 0.0025 | 0.01 | 0.035 | |
| $m_2$ | 0.008 | 0.06 (0.03-0.3) | 0.015 | $yr^{-1}$ |
| $m_3$ | 0.03 | 3 (1-3) | 0.023 | |
| $L_1$ | 1/500 | 1/1000 | 1/250 | |
| $L_2$ | 1/20 | 1/5 | 1/75 | $yr^{-1}$ |
| $L_3$ | 1/10 | ½ (½-1) | 1/100 | |
| $R_1$ | 0.85 (0.80-0.90) | 0.10 (0.10-0.30) | 0.05 (0.05-0.20) | |
| $R_2$ | 0.40 (0.30-0.50) | 0.70 (0.60-0.80) | 0.55 (0.4-0.6) | - |
| $R_3$ | 0.50 (0.40-0.60) | 0.85 (0.75-0.95) | 0.85 (0.8-0.9) | |

# Figures

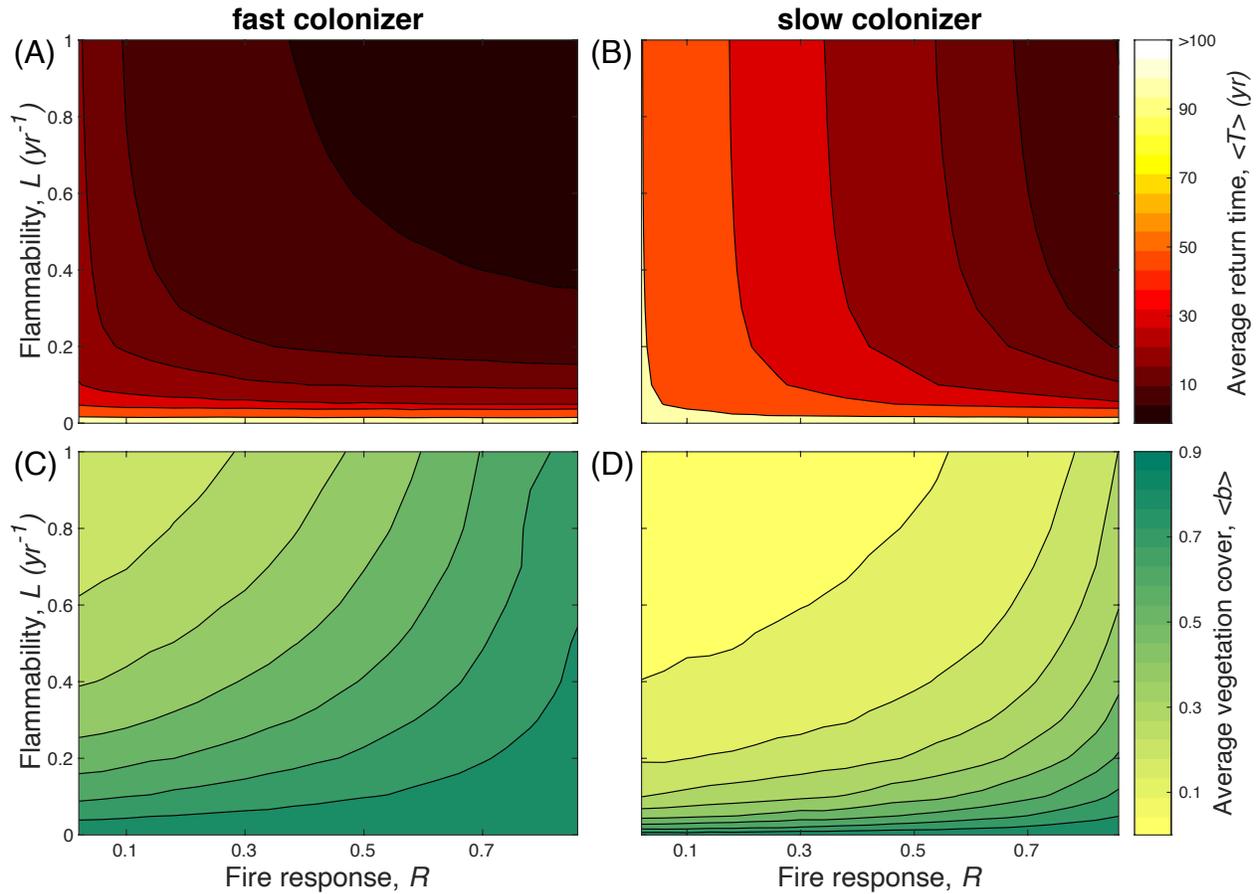

**Figure 1.** (A-B) Average fire return time ($<T>$, color scale) and (C-D) average vegetation cover ($<b>$, color scale) in the parameter plane of fire response ($R$, x-axes) and flammability ($L$, y-axes). (A-C) fast colonizer: $c=0.3\ yr^{-1}$ and $m=0.03\ yr^{-1}$. (B-D) slow colonizer: $c=0.05\ yr^{-1}$ and $m=0.005\ yr^{-1}$. The average values were computed over 50 realizations. The maximum value of $<b>$ in the absence of fire is $<b>=1-m/c$, which provides a value of 0.9 (Tilman, 1994) for both panels C and D, and across the whole parameter plane ($R, L$). The scale of the fire return time $<T>$ was arbitrarily cut at 100 yr for clarity of representation, yet values ranging up to 10,000 yr were observed at low $L$ values.

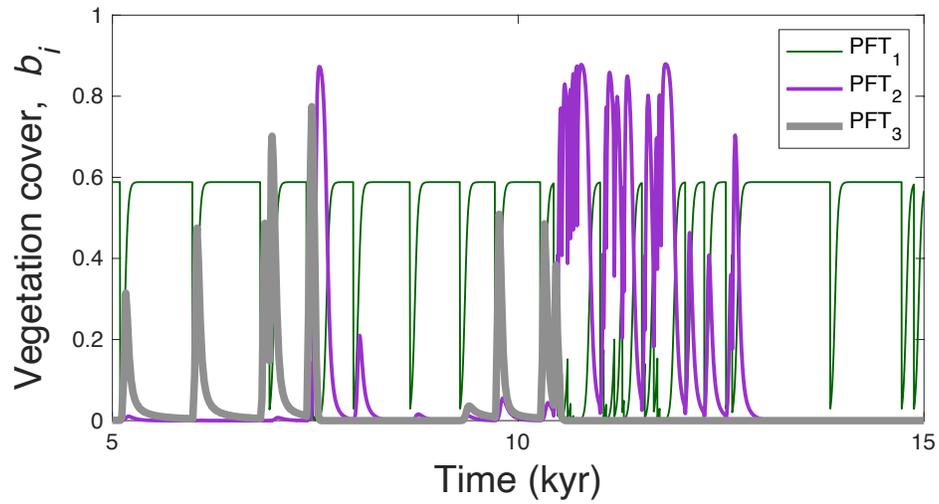

**Figure 2.** Example of a time series of fractional vegetation cover (avoider conifer PFT$_1$: green thin line; embracer conifer PFT$_2$: purple line; deciduous trees PFT$_3$: gray thick line) observed in the model for the North American boreal communities (parameter values as in Tab. 2).

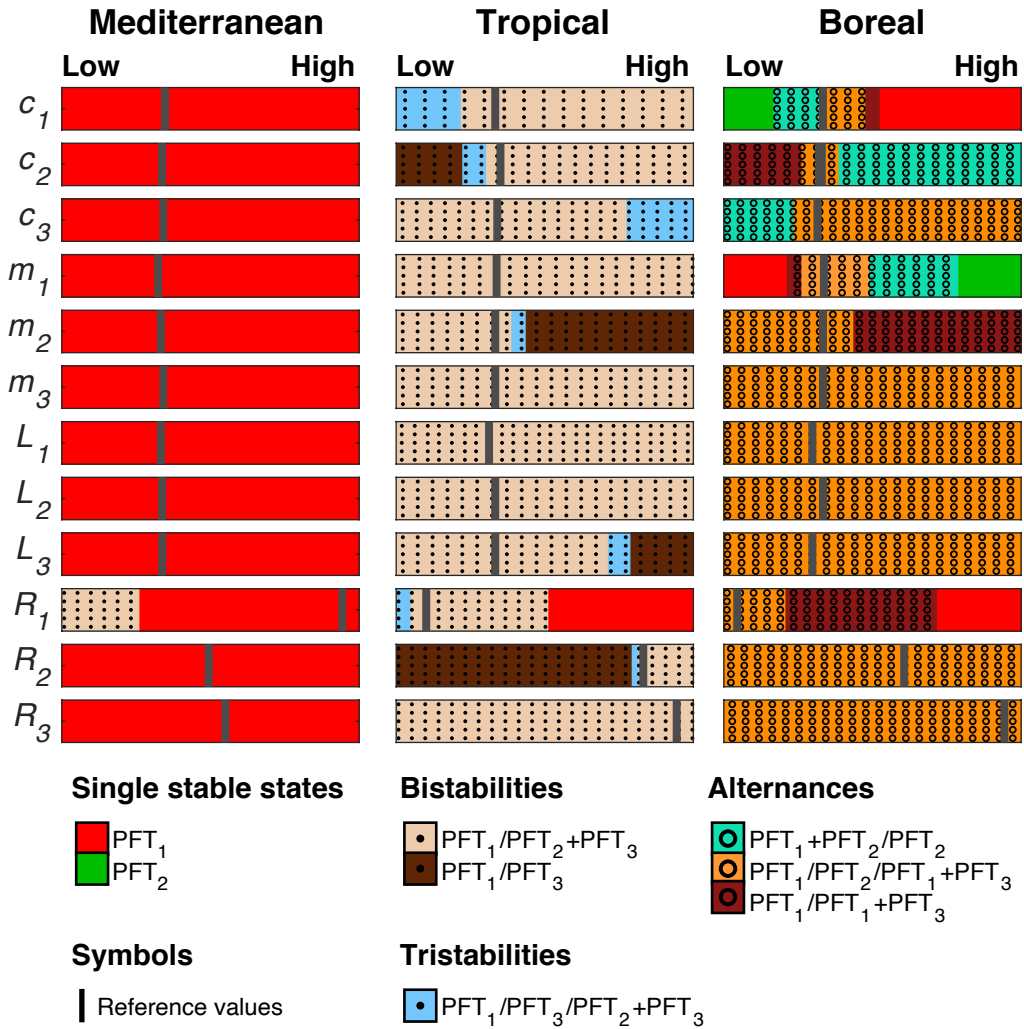

**Figure 3.** Community state maps (see color legend) observed for individual parameter variations, for the Mediterranean (left), tropical (center) and boreal (right) communities. Black vertical lines represent the reference values in Tab. 2. Parameters were individually changed from 0.5 (Low) to 2 (High) times their reference value, except for fire responses ($R_i$) that were changed between 0.01 (Low) and 0.9 (High).

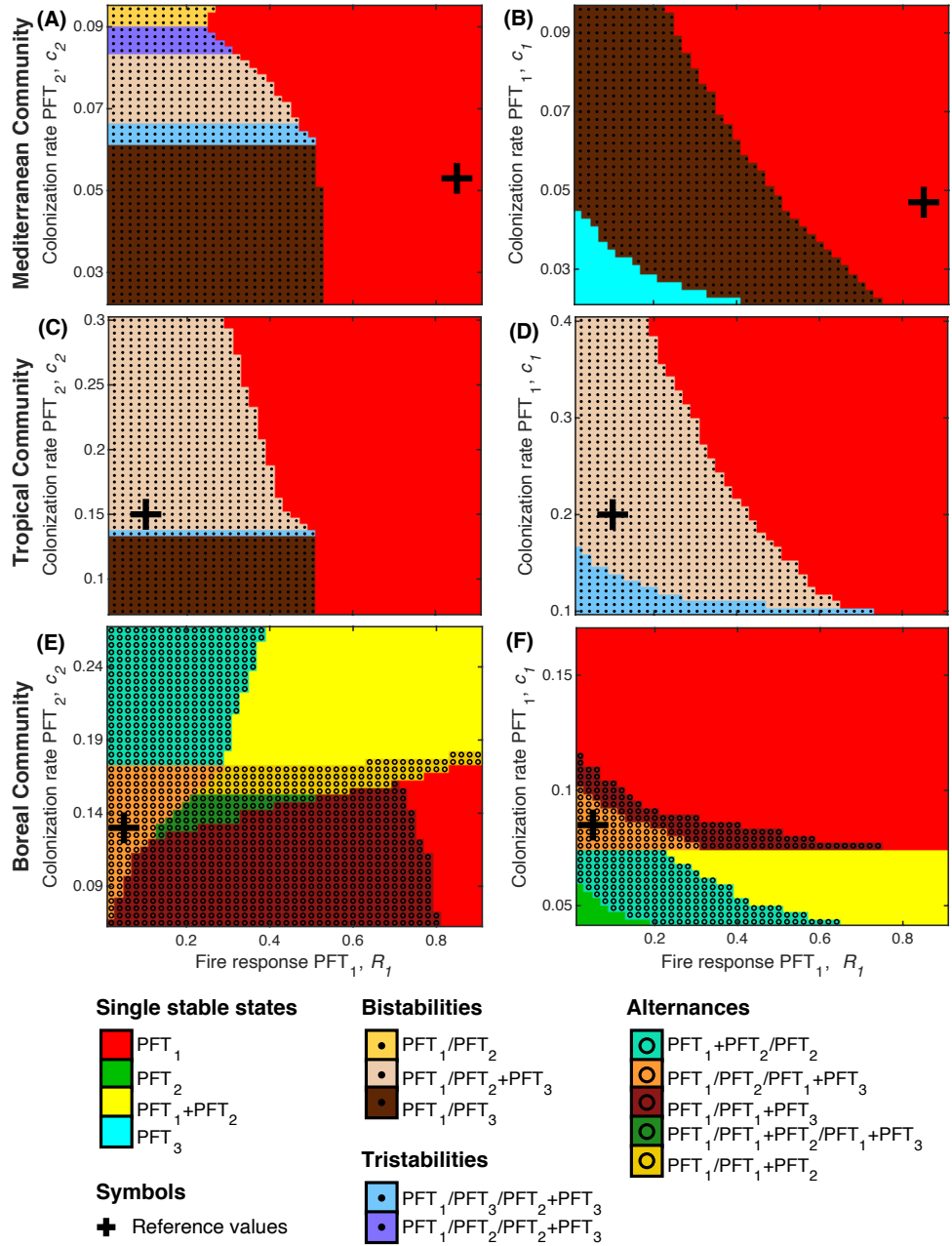

**Figure 4.** Community state maps (see color legend) observed in the parameter plane of the fire response of $PFT_1$, $R_I$ (x-axis) in combination with either the colonization rate of $PFT_2$, $c_2$ (A, C, E) or the colonization rate of $PFT_1$, $c_1$ (B, D, F) for (A-B) Mediterranean, (C-D) humid tropical and (E-F) Boreal communities. The parameter reference values (Tab. 2) are identified by the black crosses.